\documentclass[12pt,preprint]{aastex}

\shorttitle{Gamma-ray Signal from Earth-mass Dark Matter Microhalos}
\shortauthors{Ishiyama et al.}

\begin{document}

\title{Gamma-ray Signal from Earth-mass Dark Matter Microhalos}

\author{TOMOAKI \textsc{Ishiyama}\altaffilmark{1}, 
JUNICHIRO \textsc{Makino}\altaffilmark{1}, and
TOSHIKAZU \textsc{Ebisuzaki}\altaffilmark{2}}
\altaffiltext{1}{National Astronomical Observatory, Mitaka, Tokyo 181-8588, Japan ;
ishiyama@cfca.jp, makino@cfca.jp}
\altaffiltext{2}{RIKEN, 2-1, Hirosawa, Wako, Saitama 351-0198, Japan ; ebisu@postman.riken.jp 
}

\begin{abstract}
Earth-mass dark matter microhalos with size of $\sim$ 100 AUs are the
first structures formed in the universe, if the dark matter of the
Universe are made of neutralino.  Here, we report the results of
ultra-high-resolution simulations of the formation and evolution of
these microhalos. We found that microhalos have the central density
cusps of the form $\rho \propto r^{-1.5}$, much steeper than the cusps
of larger dark halos.  
The central regions of these microhalos survive
the encounters with stars 
except in very inner region of the galaxy down to the radius of 
a few hundreds pcs from the galactic center.
The annihilation signals from nearest microhalos are observed as
gamma-ray point-sources (radius less than 1'), with unusually large
proper motions of $\sim 0.2$ degree per year. 
Their surface brightnesses are $\sim$10\% of that of the galactic center.  
Their S/N ratios might be better if they are far from the galactic plane.
Luminosities of subhalos are determined only by their mass, and they are
more than one order of magnitude luminous than the estimation by \citet{Springel2008}: 
A boost factor can be larger than 1000.
Perturbations to the millisecond pulsars by
gravitational attractions of nearby earth-mass microhalos can be detected
by the observations of Parkes Pulsar Timing Array (PPTA).

\end{abstract}

\keywords{
cosmology: theory
---methods: numerical
---Galaxy: structure
---dark matter}

\section{Introduction}
In our Universe, dark matter halos evolve in the hierarchical fashion.
Smallest microhalos form first, and they merge with each other to form
larger halos. The size of the smallest microhalos is determined by the
scale of collisional damping and free streaming of dark matter particles.  
Analytical studies \citep{Zybin1999, Hofmann2001, Berezinsky2003, Green2004,
  Loeb2005, Berezinsky2008} predicted their mass to be 
$3.5 \times 10^{-9} \sim 8.4 \times 10^{-6} M_{\odot}$,
though uncertainty in theory of supersymmetry enlarges the interval further.

Early studies \citep{Berezinsky2003, Diemand2005, Berezinsky2008}
suggested that a significant fraction of microhalos born
in early universe have survived up to present time,
and they might be observed as the dominant sources of the annihilation signal.
These microhalos could enhance the annihilation signal
by a factor of 2 to 5 \citep{Berezinsky2003},
whose signature might have already been observed as electron and positron excess
[PAMELA \citep{Adriani2009}, ATIC \citep{Chang2008}, PPB-BETS \citep{Torii2008}, and
Fermi \citep{Abdo2009} ].  

\citet{Diemand2005}
simulated the formation of microhalos using $N$-body simulations.  
They argued that the density profiles were well fitted
by a single power law, $\rho(r) \propto r^{-\gamma}$,
with slope $\gamma$ in the range of 1.5 to 2 
down to the radius of $\sim 10^{-3}$ pc
and that most of
microhalos will survive against galactic tidal field and encounters
with stars.

On the other hand, \citet{Springel2008} argued that the fraction of the
mass in subhalos in the solar neighborhood was significantly lower than
that averaged over the entire halo. They assumed that the spatial
distribution of the microhalos is the same as that of subhalos with mass
$10^5M_{\odot}$, and concluded that 
the microhalos have a negligible impact on detectability.
However, it is not
clear whether their result for subhalos with a mass $10^5M_{\odot}$ can
be used to estimate the distribution of halos of $10^{-6}M_{\odot}$.

Since earth-mass subhalos contain no substructures by definition, their
central structures can be completely different from that of more massive
halos which contain many substructures.  However, there is no simulation
of earth-mass halos with sufficient resolution to study the their
central structure, so far.  \citet{Diemand2005} used mass resolution of
$1.2 \times 10^{-10}M_{\odot}$, which was too low to determine the
central structures of microhalos.

In this letter, we report the result of cosmological $N$-body
simulations with 100 times better mass resolution and 20 times better
spatial resolution compared to that used in the previous work
\citep{Diemand2005}.

\section{Initial Conditions and Numerical Method}

We performed two high resolution cosmological $N$-body simulations.
The number of particles is $1024^3$. The size of the simulation
box is 30 comoving parsecs with periodic boundary condition. 
The mass of particles is $ 9.43 \times 10^{-13} M_{\odot}$.  
For the time integration, we used the GreeM code
\citep{Ishiyama2009b, Ishiyama2009}.
We used a leapfrog integrator with shared and adaptive time steps.
The step size was determined as
$\min(\sqrt{\varepsilon/|\vec{a}_i|},\varepsilon/|\vec{v}_i|)$ 
(minimum of these two values for all particles).
We simulated them from $z=500$ to $31$.
The (plummer) softening length $\varepsilon$ was constant 
in the comoving coordinate from $z=500$ to $z=100$,
and constant in the physical coordinate ($5 \times 10^{-5}$ pc)
from $z=100$ to $z=31$.
To generate the initial particle distribution, we used the MPGRAFIC package
\citep{Prunet2009}. 
We considered two models with different initial matter power spectra.

The matter power spectrum in Model A includes the sharp cutoff
corresponding to dark matter particle with a mass of 100GeV
\citep{Green2004}.  
The power spectrum of Model B is without cutoff.
The cosmological parameters adopted are based on the concordance
cosmological model [WMAP1 \citep{Spergel2003}, $\Omega_0 = 0.268,
\Lambda_0 = 0.732, h_0 = 0.71, \sigma_8 = 0.9$].  However, we used
$\sigma_8=0.8$ to be close to recent observational ones
\citep{Spergel2007, Komatsu2009}. We did not put any thermal velocities
in initial setup of Model A to avoid unphysical density fluctuations in
small scales \citep{Colin2008}.

\begin{figure}
\centering
\includegraphics[width=13cm]{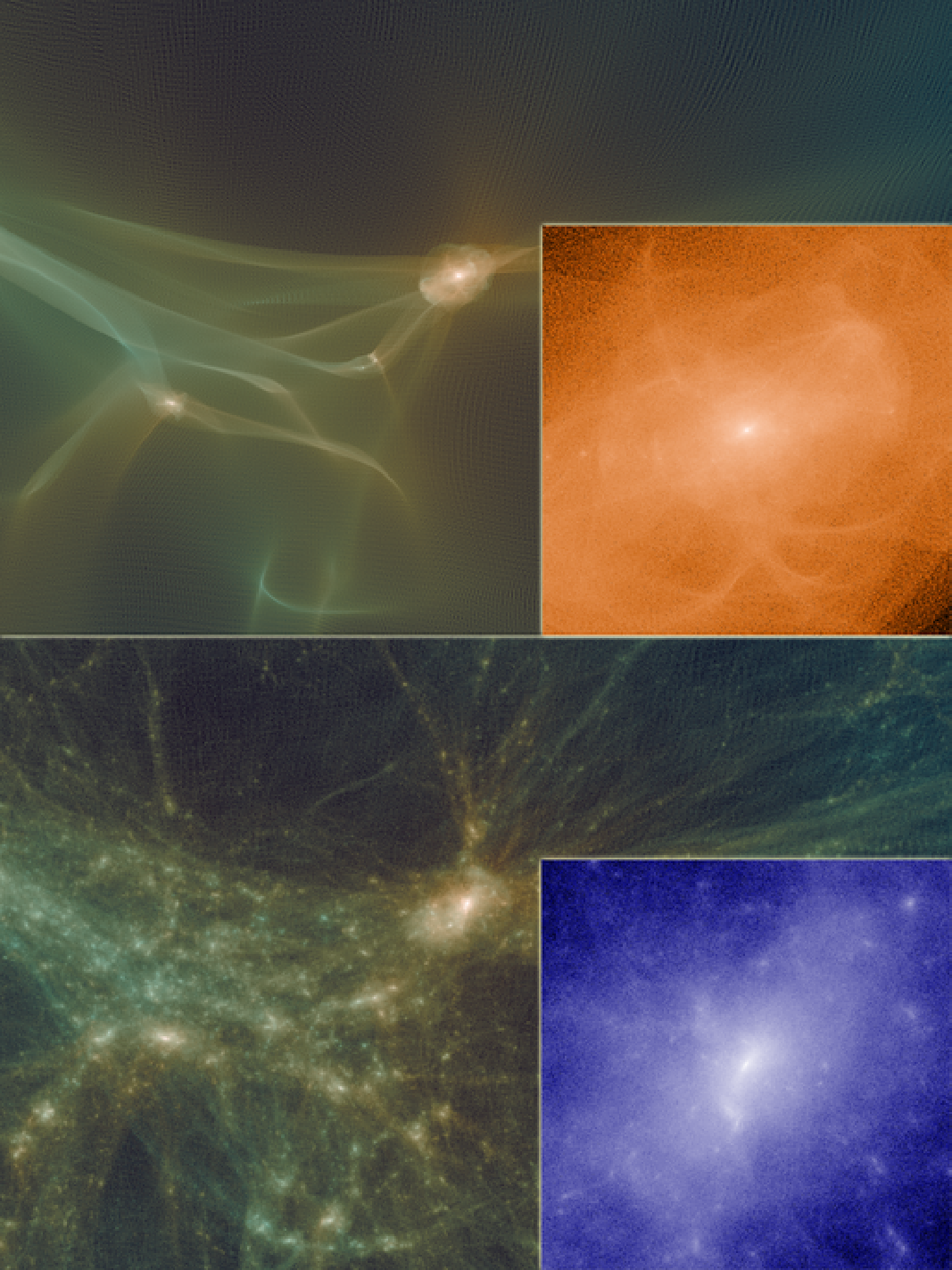}
\caption{
Top and bottom panels show the distribution of dark matter
at $z=31$ for our standard model (A) and model without free-streaming
cutoff (B).
The width of the images correspond to 12 comoving pc.
Images in the squares are enlargements of single halos.
The size of the squares is  0.6 comoving pc (4900AUs in physical units).
}
\label{fig:snapshot}
\end{figure}

Figure \ref{fig:snapshot} shows the snapshots of our simulated Universe at $z=31$.  
We can see that there are no substructures in one halo of
model A, except for caustics generated by non-linear growth of
long-wavelength fluctuations.  This structure is quite similar to what
we see in warm dark matter simulation \citep[e.g.][]{Bode2001, Gao2007},
and different from that of galaxy-sized or larger dark-matter halos
\citep{Springel2008, Ishiyama2009, Stadel2009}.  The difference comes
simply from the initial condition.  Our microhalos do not contain
smaller fluctuations inside.

\begin{figure}
\includegraphics[width=8cm]{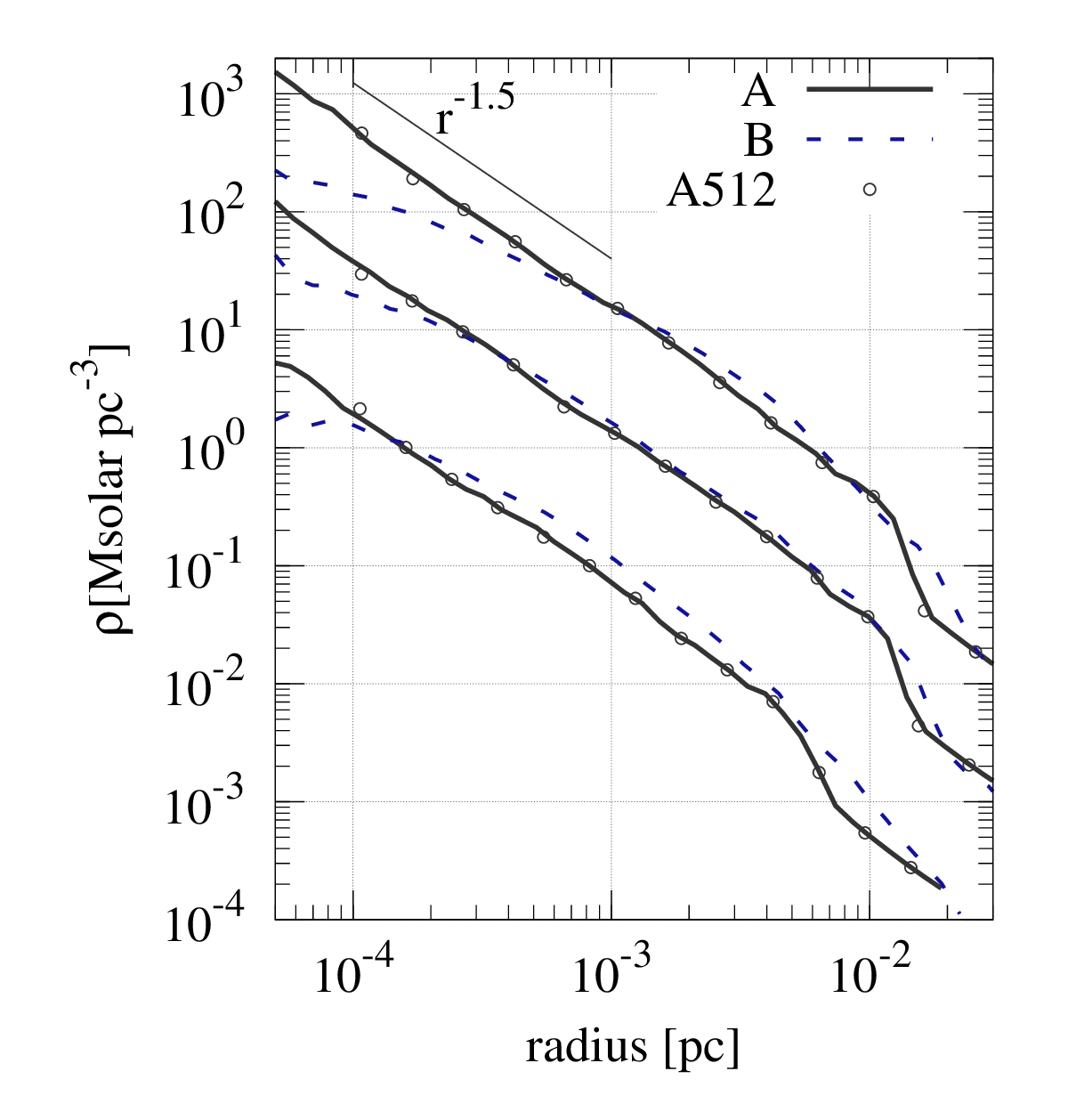}
\includegraphics[width=8cm]{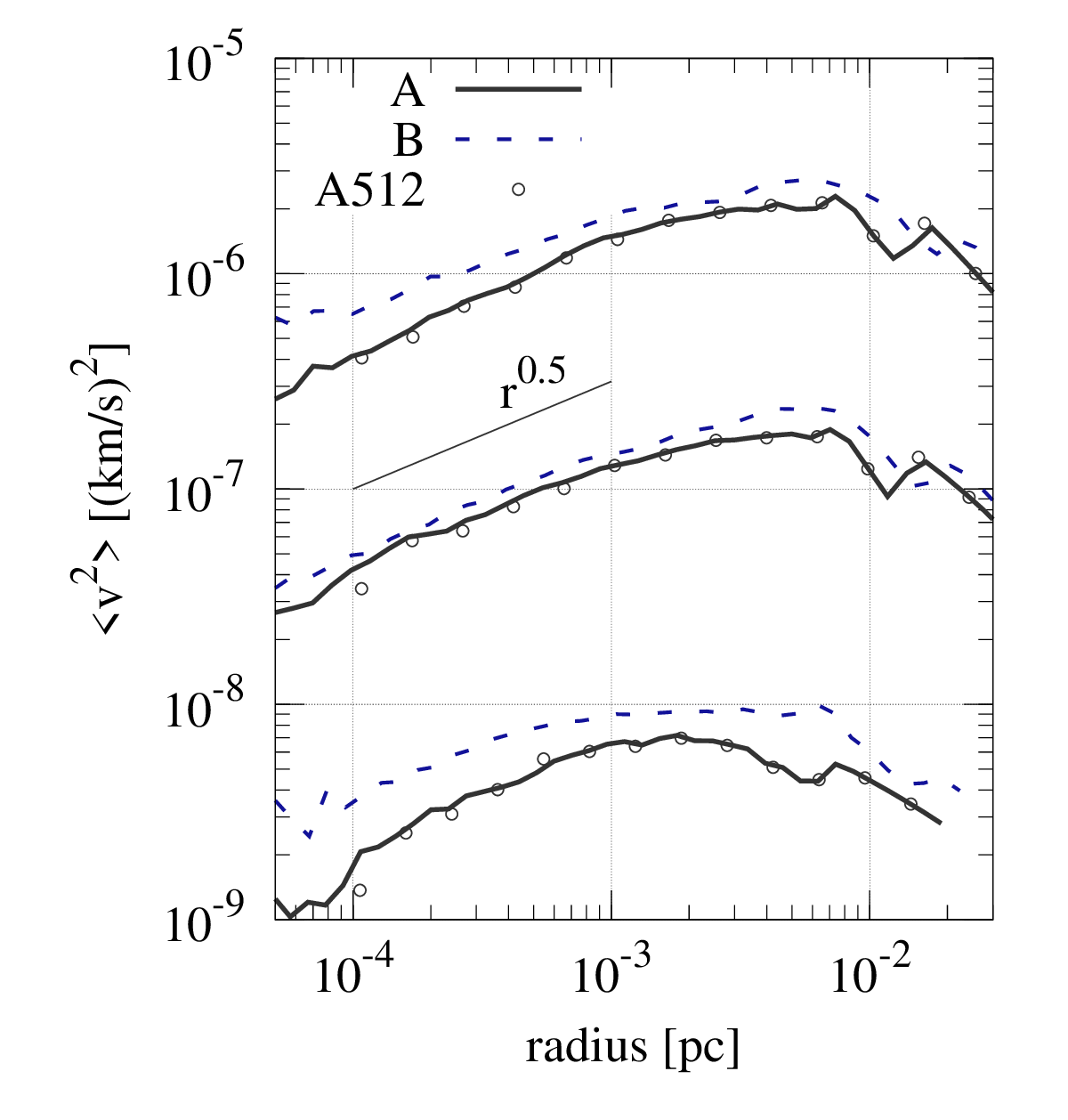}
\includegraphics[width=8cm]{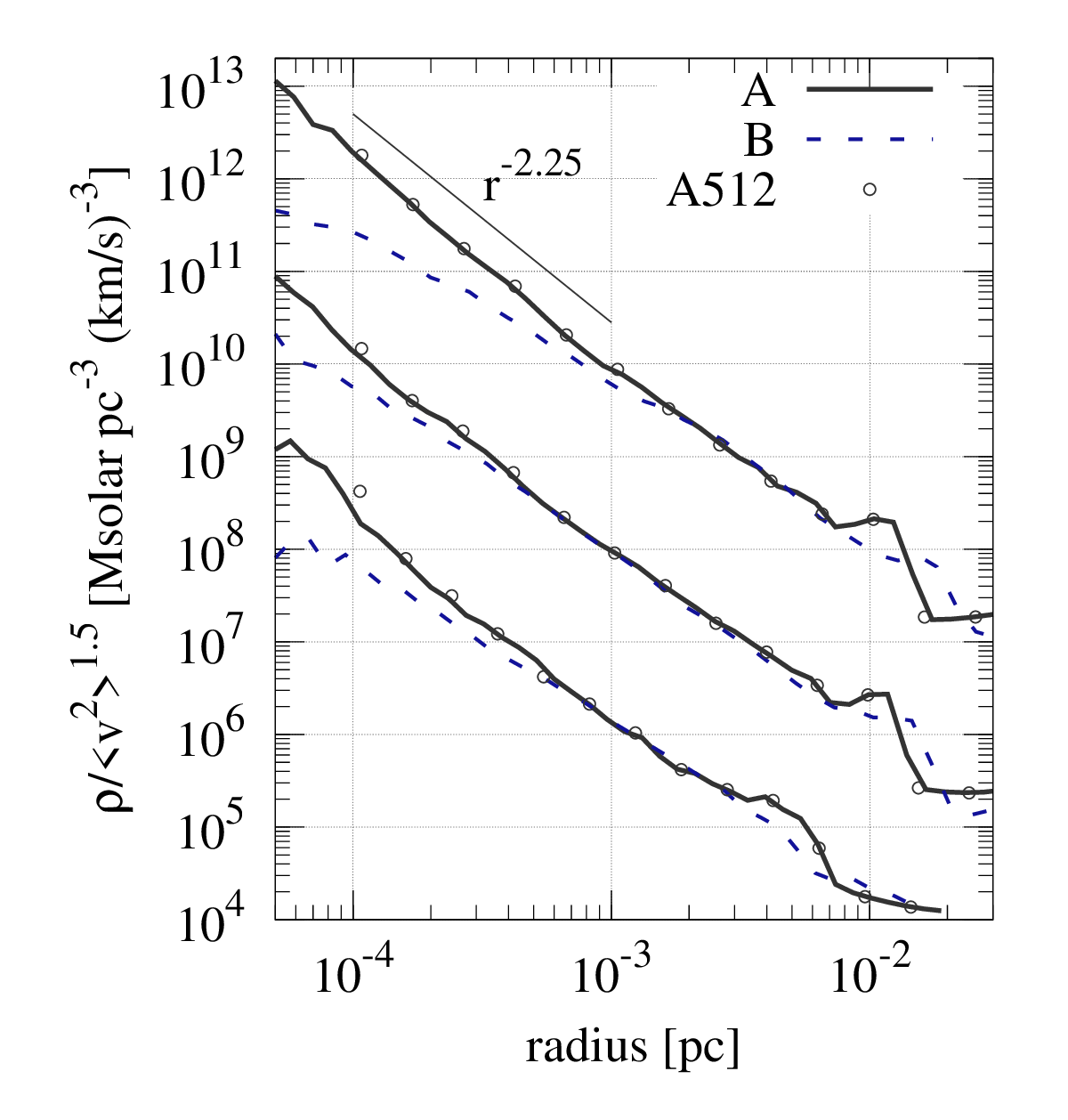}
\includegraphics[width=8cm]{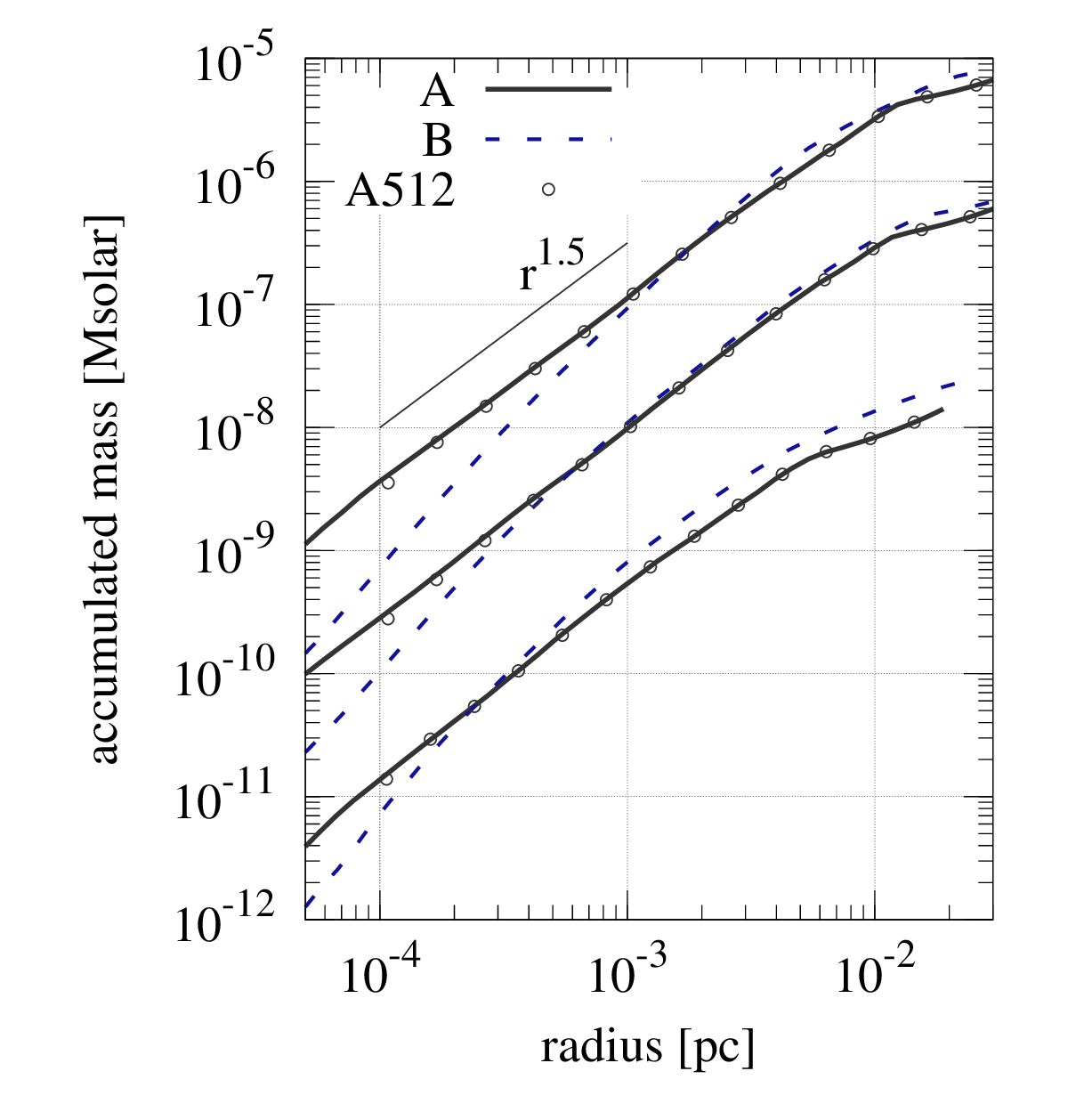}
\caption{
Radial profiles of three different microhalos of model A (black) and B
(blue) at $z=31$.  Four panels show the density, velocity dispersion, 
phase-space density, and accumulated mass. 
Circles show profiles of a low resolution simulation 
(A512: $512^3$ particles, $\rm 1.0 \times 10^{-4}$ pc softening length).
Two out of three profiles (middle and bottom) are
vertically shifted downward by 1 and 2 dex.  In the panel of phase-space
density profiles, the two profiles (middle and bottom) are shifted by 2 and 4 dex.
}
\label{fig:profile}
\end{figure}

\section{Results}

\subsection{Structures of Microhalos}

Figure \ref{fig:profile} shows the spherically-averaged structures of
three most massive halos. 
We also performed a low resolution simulation of model A for convergence check
(the number of particles was $512^3$ and the softening length was $1.0 \times 10^{-4}$ pc).
Two results are identical outside the softening radius.
The central density slope for model A
is considerably steeper than that for model B. The result of
model B is similar to that of large-scale cosmological
simulations \citep{Navarro1996, Fukushige1997, Moore1999, Fukushige2004,
Stadel2009, Navarro2010}. 
The result of model A is consistent with being a
single power law of $\rho \propto r^{-1.5}$, for quite a wide range in
the radius. 
Our result is reliable down
to $10^{-4}$ pc, where the density reaches more than $500 M_{\odot}/{\rm pc^{3}}$.

The left-lower panel of figure \ref{fig:profile} shows the rough
estimate of the phase-space density given by $\rho/\langle v^2
\rangle^{1.5}$, where $\langle v^2 \rangle$ is the velocity dispersion.
The slope is close to $-2.25$ for $r<10^{-3}$ pc.  Because of
Liouville's theorem, this phase space density cannot exceed the initial
value at the time of decoupling, which is $\sim 10^{15} {\rm M_{\odot}
pc^{-3} (km/s)^{-3}}$.  Thus, the gravitational collapse of the cusp
stops at the radius at which the phase space density reaches this
limit. This radius and density there are $r_{\rm c} \sim 10^{-5} {\rm
pc}$ and $\rho_{\rm c} \sim 2 \times 10^{4} {\rm M_{\odot}
pc^{-3}}$. We, therefore, can safely conclude the density profile of
microhalos is
\begin{equation}
\rho(r) = \rho_{\rm c} (r/r_{\rm c})^{-1.5} \quad {\rm for}\quad 
10^{-3} {\rm pc} \ge  r \ge r_{\rm c},
\label{eq:profile}
\end{equation}
and $\rho(r) \sim \rho_{\rm c}$ for $r < r_{\rm c}$.
This profile is consistent with the results of 
early studies of the collapse of self-gravitating gaseous spheres \citep{Suto1988}
and recent high resolution 
simulations of cold collapse \citep{Nipoti2006}
and warm dark matter \citep{Colin2008}.

\begin{figure}[t]
\includegraphics[width=12cm]{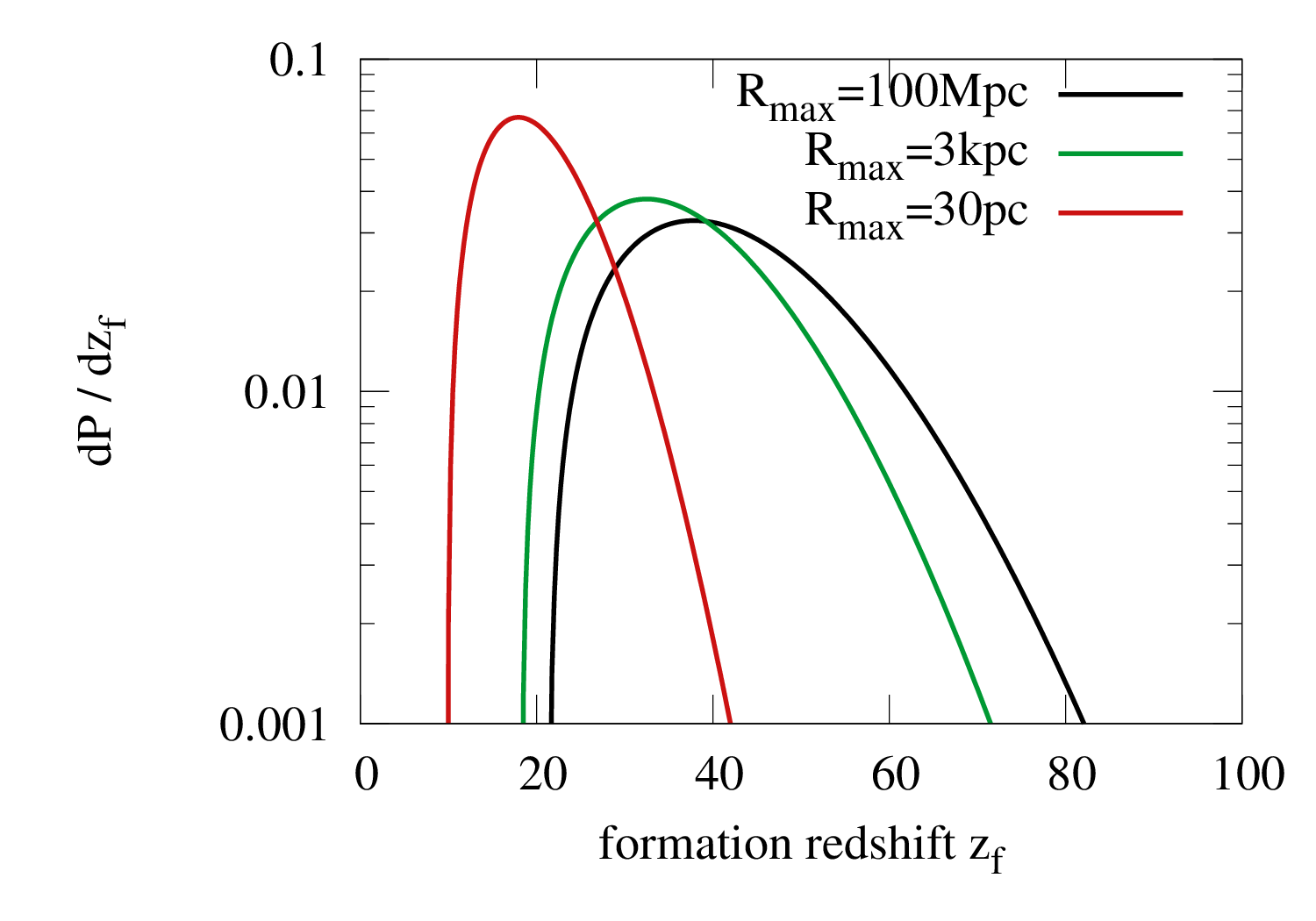}
\caption{
The differential probability distribution of the microhalo formation redshift 
predicted by Press-Schechter theory.
The longest wavelength correspond to 100Mpc (black), 3kpc (green) and 30pc (red). 
These value corresponds to the box length of the simulations 
by \citet{Diemand2005} (3kpc) and this work (30pc).
}
\label{fig:psfunc}
\end{figure}

The average density of these earth-mass halos is 
affected by density fluctuations of larger scales. 
However, our simulations do not contain large scale fluctuations.
As seen in Figure \ref{fig:psfunc}, 
Press-Schechter \citep{Press1974} theory predicts that 
the average formation epoch of microhalos is earlier
if we take into account larger scale fluctuations. 
Since the average density of a microhalo 
should be proportional to the cube of its formation redshift, 
most microhalos should have higher average density than these in our calculation.
For simplicity, we use the profile of
equation (\ref{eq:profile}) to estimate the chance of survival of these halos.

\subsection{Tidal Cutoff and Encounters with Stars}\label{sec:encounter}

The tidal cutoff radius of microhalo with the density profile of 
equation (\ref{eq:profile}) due to galactic tide is expressed as 
$r_{\rm max} = 0.082(R/10{\rm kpc})^{4/3} {\rm pc}$, 
where $R$ is the distance from the galactic center, if we 
assume an isothermal halo with the rotation velocity of 220
km/s. We can see that these halos can easily survive at the distance
1kpc, and even at the distance 0.1kpc, the central $10^{-4}$ pc
of the halo would still survive. 

Can encounters with stars destroy the internal cusps of
microhalos \citep{Berezinsky2003, Berezinsky2006, Angus2007, Goerdt2007, Zhao2007,
  Green2007, Schneider2010}?  Since the surface number density of disk stars is about
100 $\rm pc^{-2}$ at the solar neighborhood, the closest distance a
halo with the orbital period of the order of 100 Myrs can approach to
a star in the Hubble time is $10^{-2}$ pc. 

\begin{figure}
\centering
\includegraphics[width=5.4cm]{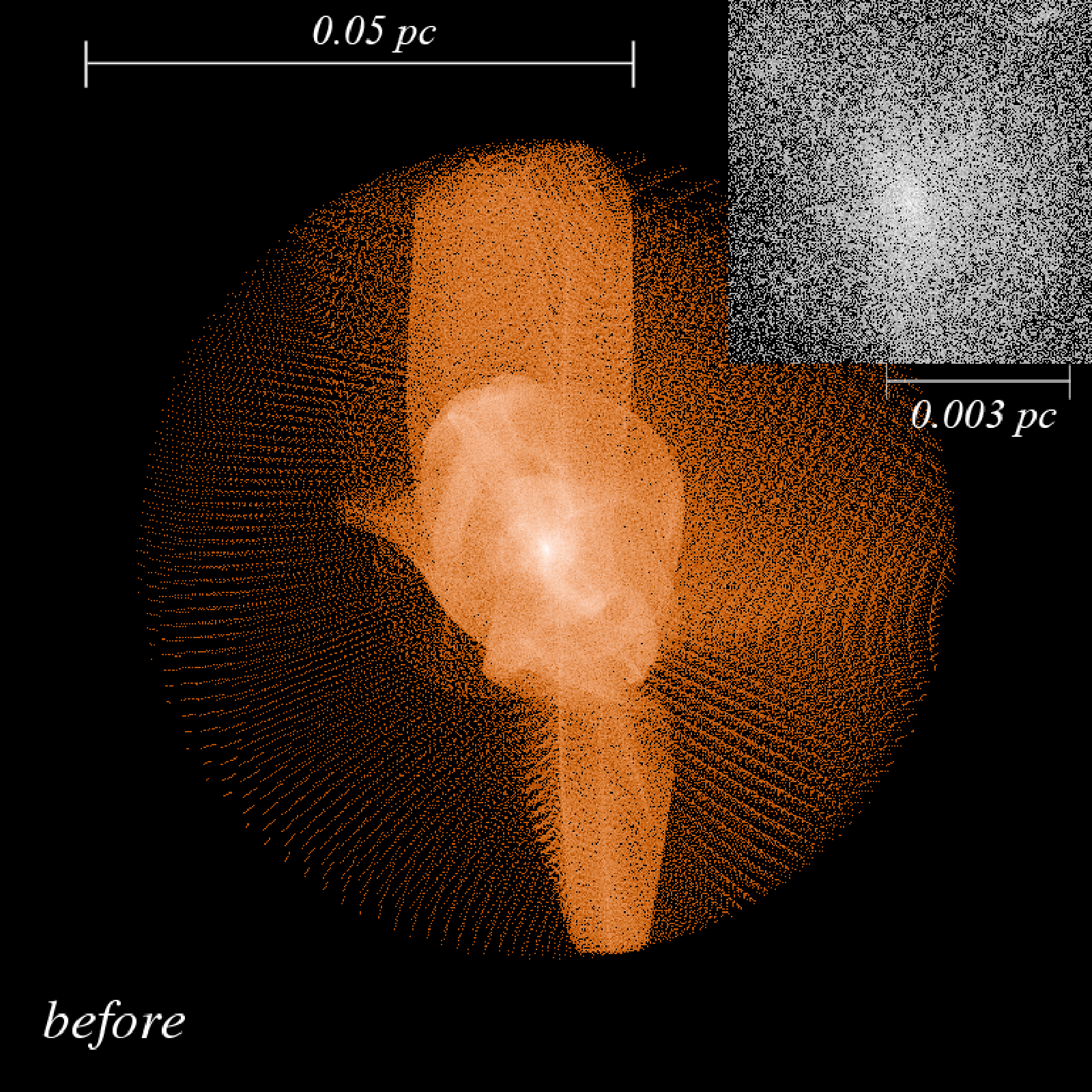}
\includegraphics[width=5.4cm]{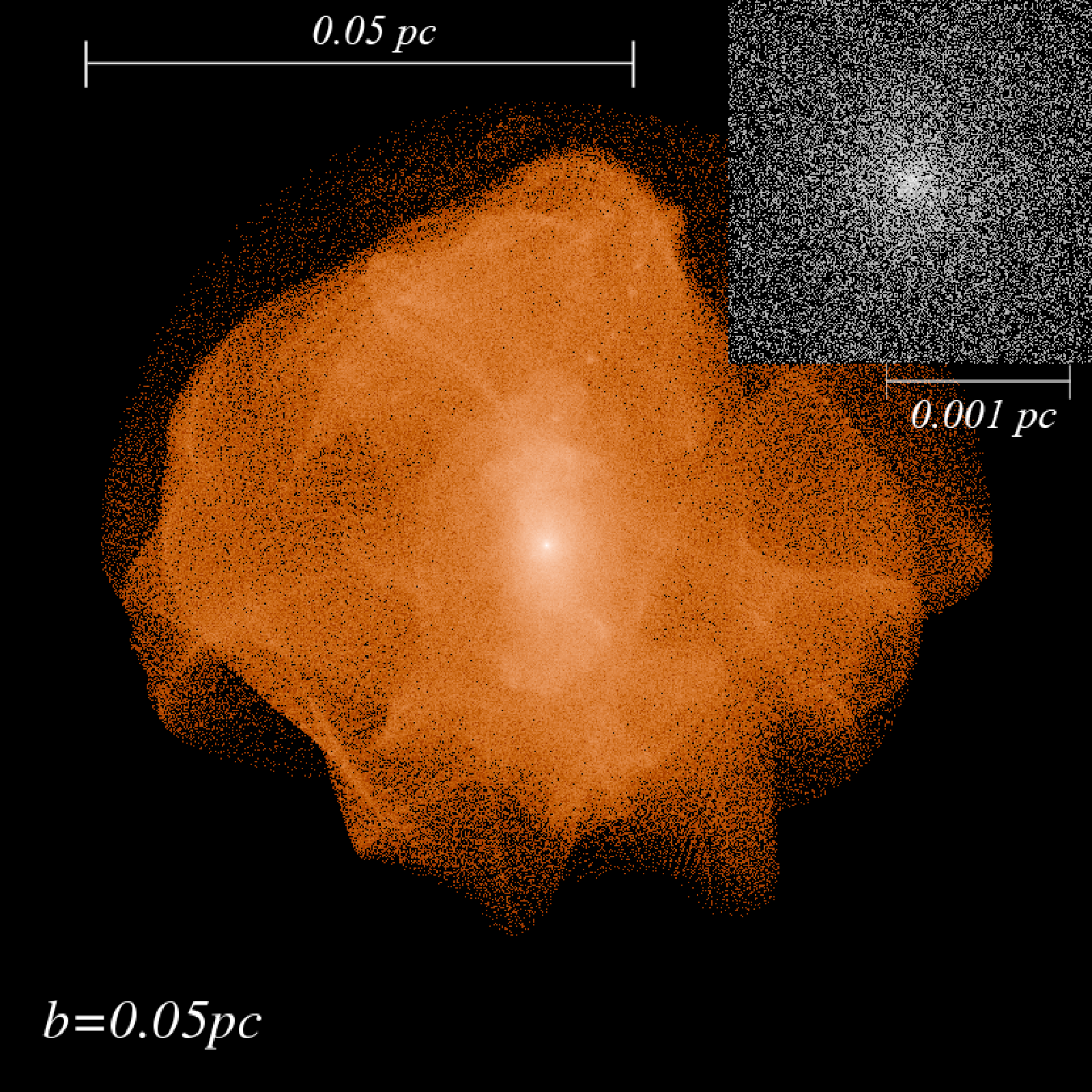}
\includegraphics[width=5.4cm]{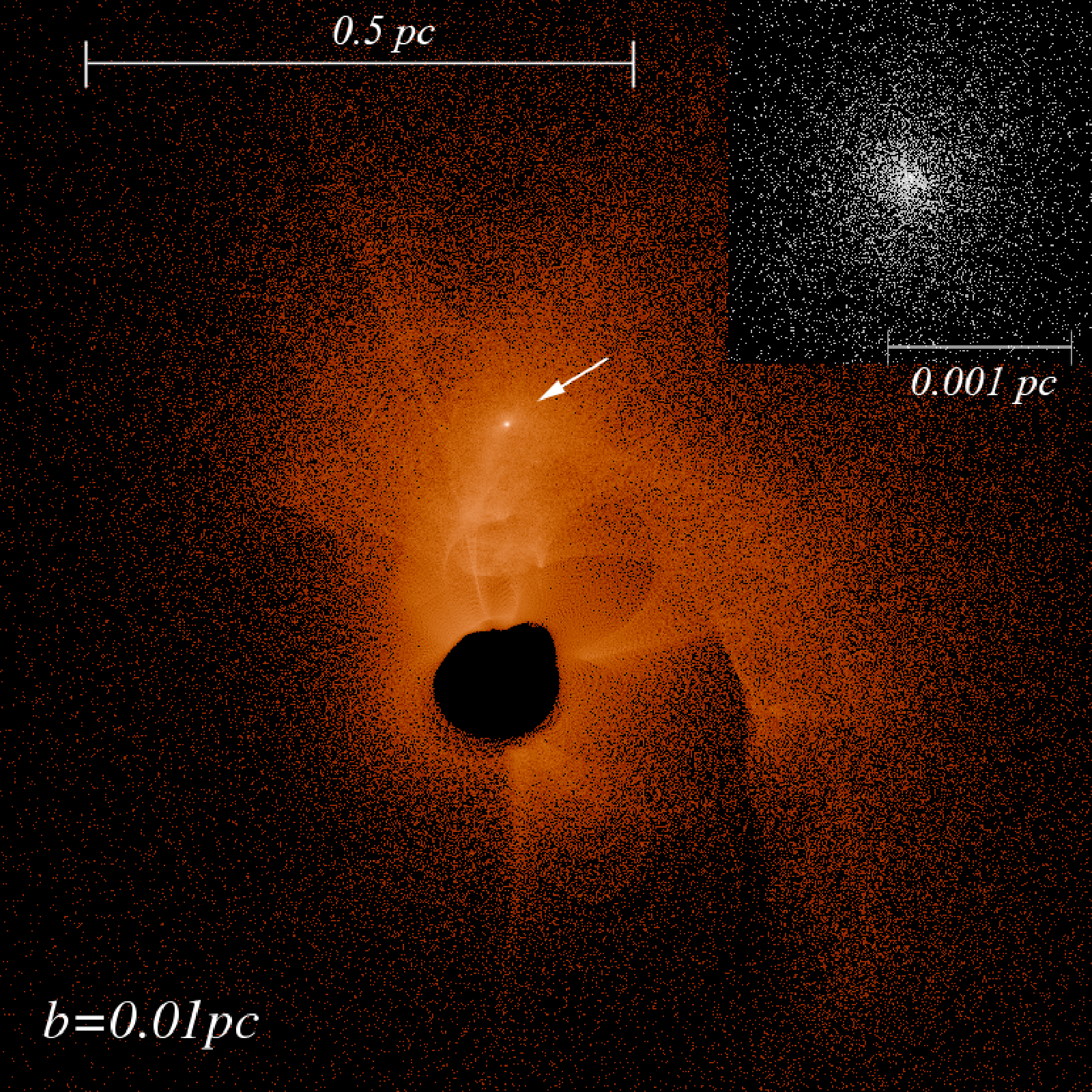}
\includegraphics[width=5.4cm]{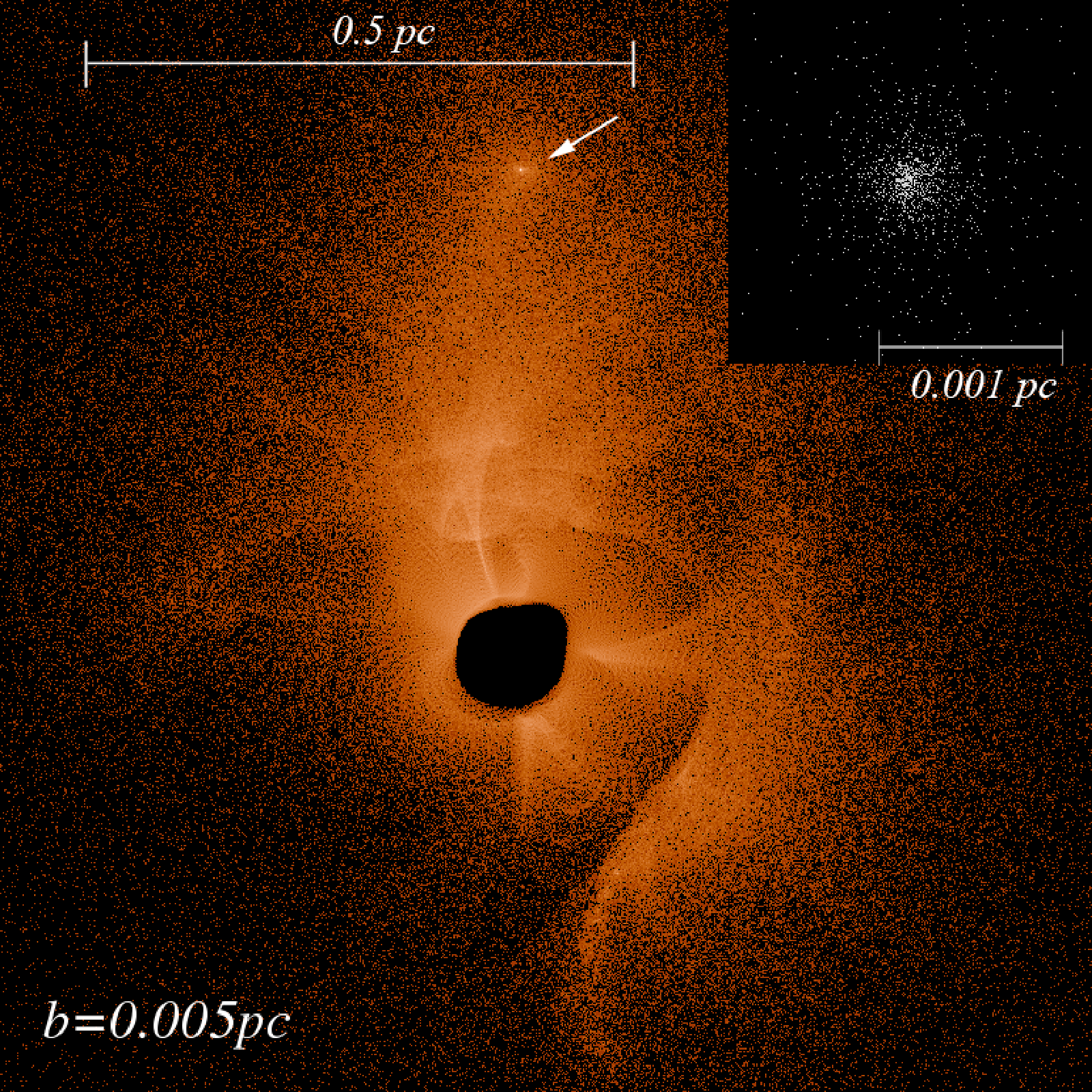}
\includegraphics[width=5.4cm]{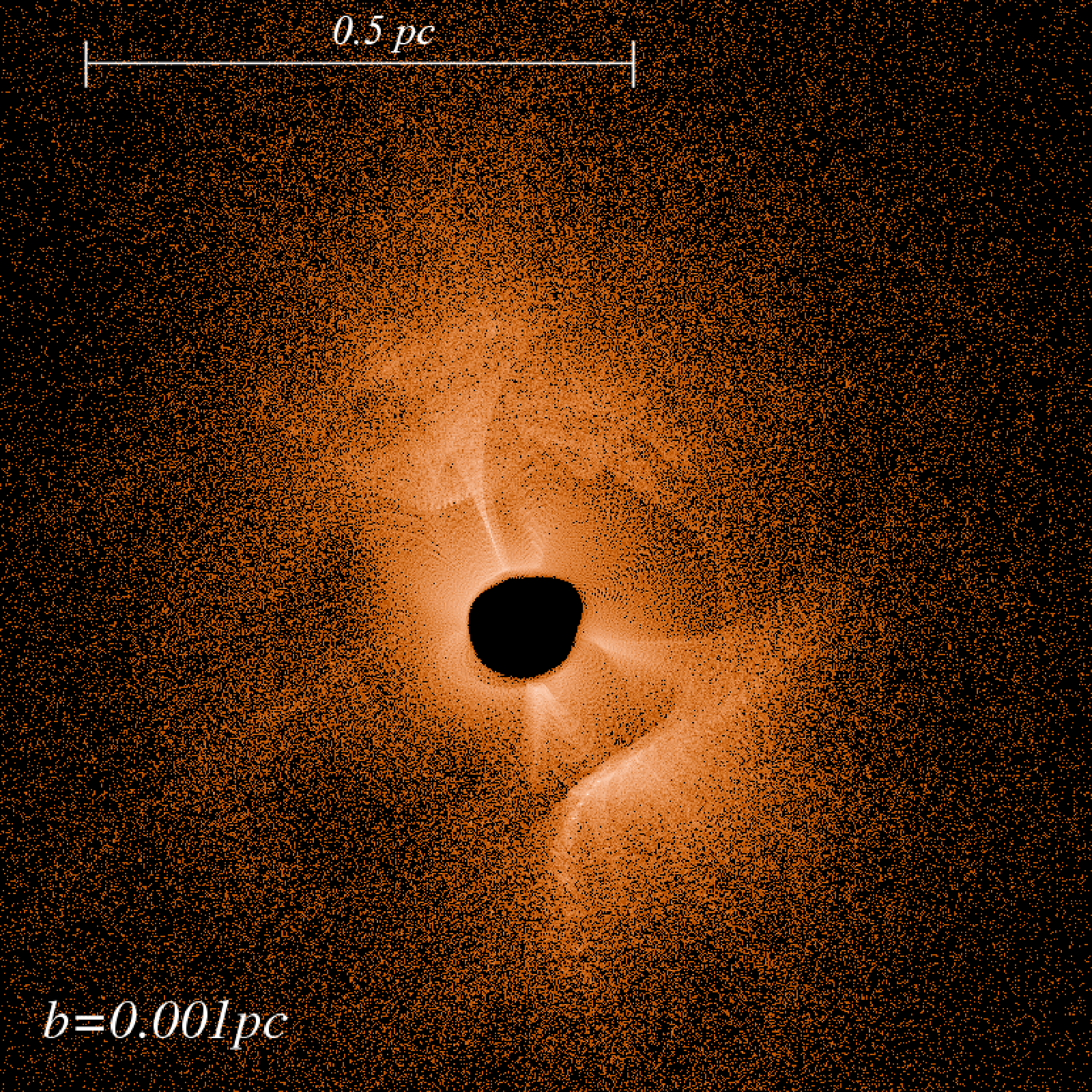}
\includegraphics[width=5.4cm]{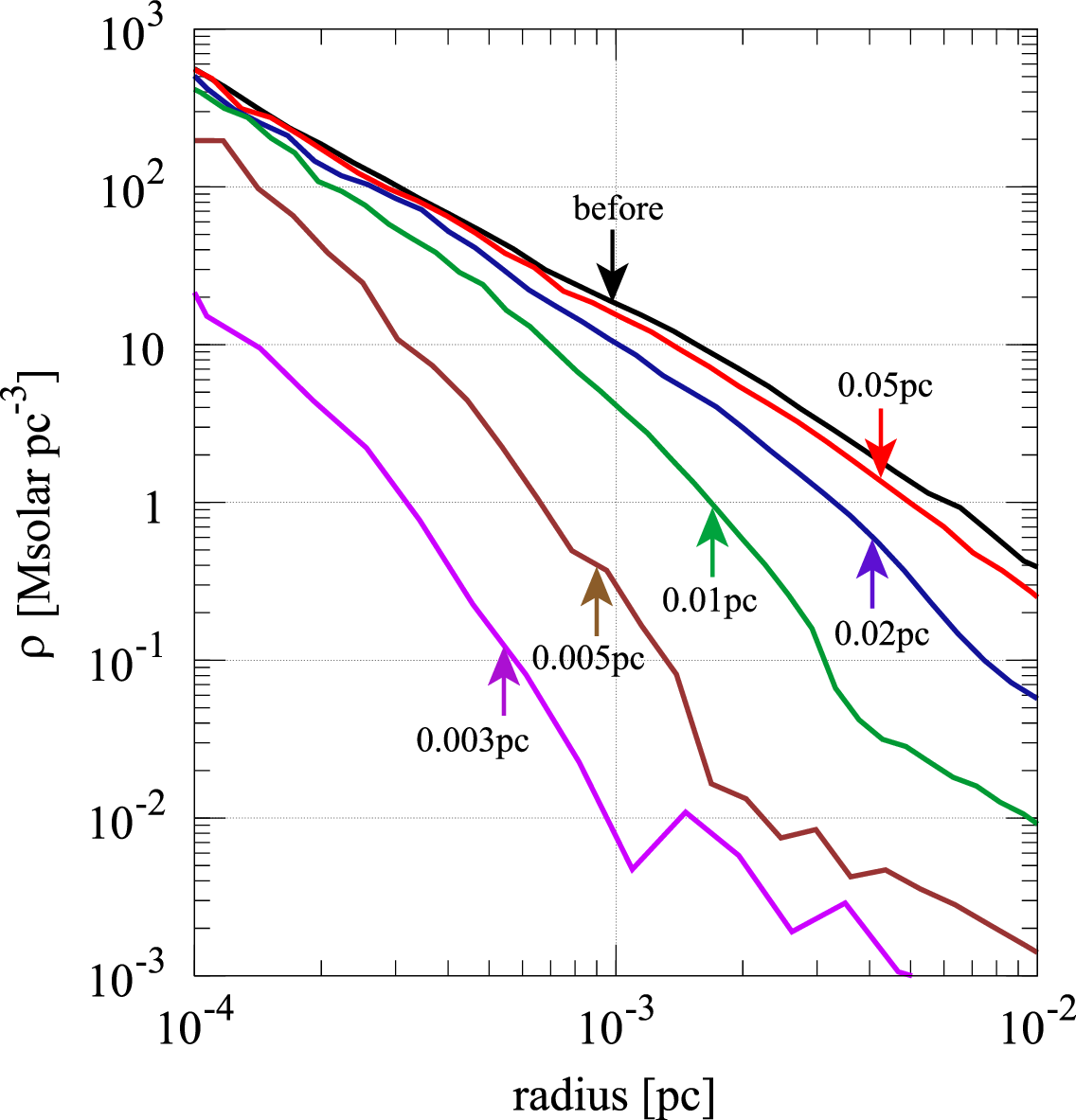}
\caption{
Snapshots and radial density profiles after encounters with stars.
Panel a (top-left) shows the snapshot  of a halo in model A at
$z=31$. Only particles within the radius of 0.037 pc from the center of
the halo are shown.   Panel b (top-middle), c (top-right), 
d (bottom-left), and e (bottom-middle)
show the distribution of particles at 27 Myrs after the encounter with
a solar-mass star with the impact parameters $5 \times 10^{-2}$pc,
$1 \times 10^{-2}$pc, $5 \times 10^{-3}$pc, and $1 \times 10^{-3}$pc. 
The black regions in the center are caused by the initial
outer cutoff of the particle distribution and are not real. 
The upper right box in each panel shows the central region of the microhalo.
The central regions are located at the center (Panel a and b), and
the positions indicated by arrows (Panel c and d).
In panel e, the central region is destroyed completely.
Panel f (bottom-right) shows the density profiles after the
encounter. 
}
\label{fig:encounter}
\end{figure}

Figure \ref{fig:encounter} shows the result of encounter 
with a $1 M_{\odot}$ star moving at $\rm 200kms^{-1}$.  
We selected the most massive microhalo at $z=31$ from the simulation of model A 
(the cutoff radius is two times virial value), 
and added velocities on each particle according to the impulsive approximation.
Then, we simulated the evolution of the microhalo for 27Myr after the encounter.
We can see that the regions of radius  $\sim 10^{-3}$ pc survives
after encounters with the impact parameters of 
$0.02$ pcs. Theoretically, the cutoff radius due to
encounter for the power-law cusp with slope $-1.5$ is given by $r
\propto b^{8/11}$, where $b$ is the impact parameter. Therefore,
the complete disruption requires an encounter with $b=5\times 10^{-5}{\rm
  pc}$. Such close encounters are expected only at the very central
regions of the galaxy (20pc or less from the center).

If we extrapolate the subhalo mass function $dn/dm \propto m^{-2}$
\citep{Berezinsky2003} to the mass of microhalos, the number of
microhalos in our Galaxy is $\sim 5 \times 10^{16}$ (we assumed the mass
function $dn/dm \sim 5 \times 10^{10} m^{-2}$).  The number density of
the microhalos near the solar neighborhood is $\rm \sim 500 pc^{-3}$, if
the distribution of the microhalos is the same as that of the background
dark matter.  This extrapolation is justified by the fact that the
microhalos survive under tides from our Milky Way Galaxy.  Note that the
above estimate takes into account the effect of merging between small
halos and accretion to somewhat larger halo since we extrapolated the
mass function obtained by $N$-body simulation.

\subsection{The Gamma-ray Luminosity of a Microhalo}

The gamma-ray flux of a microhalo is given by
\begin{eqnarray}
F_\gamma &\sim&  \frac{{N_\gamma} \langle \sigma v \rangle}{2m_{\chi}^2}
\frac{1}{4\pi d^2} 
\int_0^{r_{\rm max}} \rho^2 dV \\
&\sim& \frac{{N_\gamma} \langle \sigma v \rangle}{2m_{\chi}^2}
d^{-2} 
\ln(r_{\rm max}/r_{\rm c}) \rho_{\rm c}^2 r_{\rm c}^3 , \\
&\sim& 9.2 \times 10^{-12} \frac{N_\gamma}{30} 
\frac{\langle \sigma v \rangle}{\rm 3 \times 10^{-26}cm^3s^{-1}}
\left( \frac{m_{\chi}}{\rm 100GeV} \right)^{-2}
\left( \frac{\rm 0.2pc}{d} \right)^2 \label{eq:gammarayflux2} \\
&\quad&
\left(\rm \frac{\rho_{c}}{2\times10^4M_{\odot}pc^{-3}} \right)^2
\left(\rm \frac{r_{c}}{10^{-5}pc} \right)^3 
\left( 4.61 + \ln{\rm \frac{r_{max}/10^{-3}pc}{r_{c}/10^{-5}pc}} \right)
\quad {\rm photons \cdot cm^{-2} \cdot s^{-1}}, \nonumber
\end{eqnarray}
where $N_\gamma$ is the number of emitted photon per annihilation, 
$m_\chi$ is the mass of dark matter particle, 
$\langle \sigma v \rangle$ is the interaction cross section of dark matter,
and $r_{\rm max}$ is the outer cutoff radius of a microhalo at which the slope
of the density becomes steeper than $-1.5$.

Mass loss due to the galactic tide or encounters with stars
changes  $r_{\rm max}$ of microhalos, but does not affect
$\rho_{\rm c}$ or $r_{\rm c}$ except in the cases of extremely
close encounters,  as we can see in figure 3.
Therefore, they have relatively minor effect on the gamma-ray luminosity.  
After 99\% of the mass is lost from a microhalo, it still
retains nearly 50\% of the luminosity. 
Thus, we can assume that
the tidal stripping has practically no effect on the gamma-ray
luminosity of our earth-mass halos. 
These results supports the predictions of early analytical studies \citep{Berezinsky2008}.

Here, we consider how the gamma-ray flux of 
the microhalo depends on its formation epoch. 
The average density of a halo $\rho_{\rm ave}$ reflects the 
cosmic density at its formation time  \citep{Bullock2001}.
Since the cosmic density is proportional to $(1+z_{\rm f})^3$, 
$\rho_{\rm ave} \propto \left( \frac{1+{z_{\rm f}}}{1+{z_{\rm 0}}} \right)^3$, 
where $z_{\rm f}$ and $z_{\rm 0}$ is the formation redshift and 
the typical formation redshift of microhalos in our simulations. 
From the conservation of the mass, we can derive $\rm r_{max} \propto \rho_{\rm ave}^{-1/3}$.
We can rewrite $\rho_c^2 r_c^3$ in equation (\ref{eq:gammarayflux2}) as 
$\rho_{\rm ave}^2 r_{\rm max}^3$ using equation ({\ref{eq:profile}}). 
Therefore, 
\begin{eqnarray}
F_\gamma (z_{\rm f}) \sim 
\left( \frac{1+{z_{\rm f}}}{1+{z_{\rm 0}}} \right)^3 F_\gamma. \label{eq:gammarayflux3}
\end{eqnarray}
Here, we drop the dependency of the formation redshift 
in the logarithmic term in equation (\ref{eq:gammarayflux2}), 
since it is rather weak ($r_c \propto r_{\rm max}^{-1/3}$).

The average boost factor due to the formation epoch is determined as
\begin{eqnarray}
\int \frac{dP(z_{\rm f})}{dz_{\rm f}} \frac{F_\gamma (z_{\rm f})}{F_\gamma(40)}  dz_{\rm f} , 
\end{eqnarray}
where $P(z_{\rm f})$ is the distribution function of the formation epoch.
For the PS function in Figure \ref{fig:psfunc}, 
its value is $\sim$ 1.6.
We apply this boost factor to estimate the gamma-ray luminosity.

\section{Discussions and Summary}

\subsection{Gamma-ray Signal from Microhalos}

There are many works on whether or not subhalos and microhalos can be
observed via annihilation gamma-ray \citep[e.g.][]{Oda2005,
Koushiappas2006, Ando2008, Lee2009, Kamionkowski2010, Schneider2010}.
We made the all sky map of the gamma-ray annihilation signal in our
Galaxy based on the new profile of the microhalo.  The observer locates
at 8.5 kpc from the center of the halo along its long axis.  We consider
the emissions from microhalos only.  We assume the spatial distribution
of microhalos follows the mass distribution of a galaxy sized LCDM halo
selected from our previous simulation \citep{Ishiyama2009b}.  Thus, our
all-sky map naturally include all subhalos with the mass larger than
numerical limit ($\sim 10^8 M_{\odot}$).  In addition, destruction due
to encounters with stars are taken into account.  We assumed a
exponential disk with the surface density, $\Sigma(r) \sim 1000
\exp(-r/3 \rm kpc) pc^{-2}$, where $r$ is the distance from the galactic
center.  From the position of an ensemble of microhalos, the impact
parameter $b$ of encounters with stars are calculated.  Then, we use the
cutoff radius $r_{\rm cut} = 0.017 b^{8/11} \rm pc$ and reduce the
gamma-ray luminosity.

In order to include the luminosities of nearby microhalos, 
we placed randomly 2000 microhalos less than 1pc from the observer.
The distribution of formation redshift and the gamma-ray flux 
of microhalos are given by Figure \ref{fig:psfunc} 
and equation (\ref{eq:gammarayflux3}).

\begin{figure}
\centering
\includegraphics[width=11cm,angle=-90]{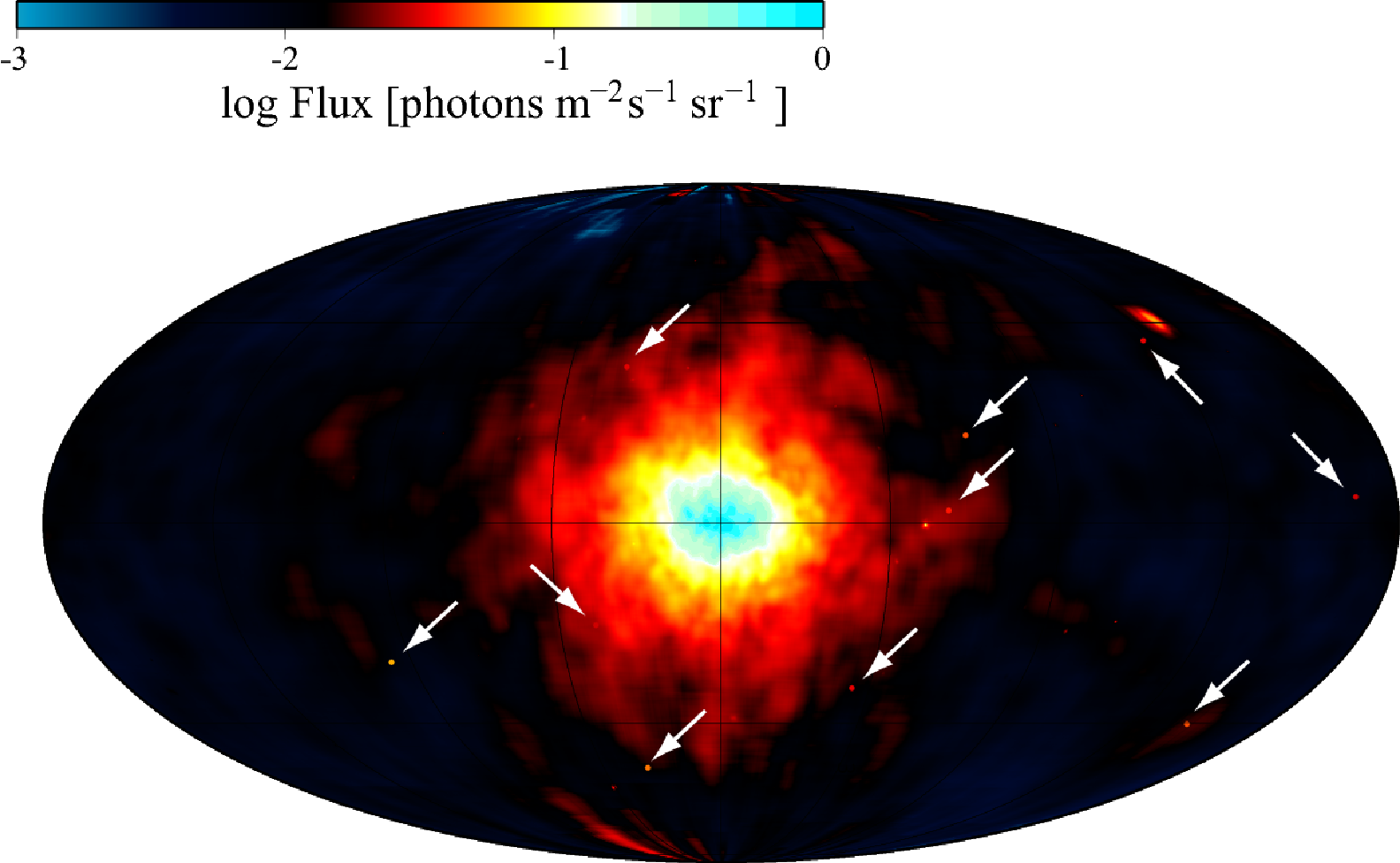}
\caption{
All sky map of the gamma-ray annihilation signal of our Galaxy.
Pointlike sources indicated by arrows show 
the contributions from microhalos less than 1pc from the observer 
and with flux larger than $\rm 0.03 \thinspace photons \cdot m^{-2} \cdot s^{-1} \cdot sr^{-1}$.
Here, we assume that the number of photons per annihilation is 
$N_{\rm \gamma} = 30$, 
the interaction cross section of dark matter is
$\langle \sigma v \rangle = 3.0 \times 10^{-26} {\rm cm^{3} s^{-1}}$,
and the mass of dark matter particle is $m_\chi = 100 {\rm GeV}$.
The flux of the integrated emission of microhalos 
in the galactic center is 
$\rm 1.40 \thinspace photons \cdot m^{-2} \cdot s^{-1} \cdot sr^{-1}$.
The flux of the most brightest microhalo is 
$\rm 0.069 \thinspace photons \cdot m^{-2} \cdot s^{-1} \cdot sr^{-1}$.
Flux needed for a 10$\sigma$ detection for Fermi is 
$\rm \sim 0.4 \times 10^{-6} photons \cdot cm^{-2} \cdot s^{-1}$ 
for low Galactic latitude and is 
$\rm \sim 0.1 \times 10^{-6} photons \cdot cm^{-2} \cdot s^{-1}$ 
for high Galactic latitude \citep{Fermi2009}.
The angular resolution of Fermi is about 0.1 degree for energies above 1GeV.
}
\label{fig:gamma-ray}
\end{figure}

Figure \ref{fig:gamma-ray} shows the all sky map.  Many nearby
microhalos are observed as pointlike sources, since their angular size
is around 1'.  Therefore, the nearby microhalos are promising sources of
gamma-ray signal of the annihilation of dark matter particles.  Their
distance is around $10^4 {\rm AU}$ and velocity is of the order of $200
{\rm km/s}$.  The proper motion is as large as 0.2 deg/year.  They might
be too dim to be observed by Fermi.  Sommerfeld effect could boost the
interaction cross section of dark matter, and enhance by orders of
magnitude the gamma-ray luminosity \citep{Kuhlen2009}.  These microhalos
might be good targets of next generation Cherenkov telescope, such as
CTA\footnote{http://www.cta-observatory.org/}.

Bright microhalos enhance luminosities of all subhalos (including known
Milky Way satellite) since subhalos contain a number of microhalos.  The
brightest gamma-ray source is the integrated emission of microhalos in
the galactic center, but it is much more extended compared to the result
of previous studies \citep{Springel2008}.  These sources may be also
good targets for observation.  However, baryons would have significant
effect on the distribution of dark matter at the center of the galaxy.
If baryons disrupt the
central cusp of the main halo \citep{Mashchenko2006, Mashchenko2008,
Governato2010}, gamma-ray flux from the galactic center might be much
smaller.  Observational data suggests that dwarf galaxies have constant
density cores \citep[e.g.][]{Gentile2004}.

\subsection{Perturbations on Millisecond Pulsars}
Pulsar timing measurements might be used to detect microhalos. Since
many of known millisecond pulsars (MSPs) are in the direction of the
galactic center, the number density of microhalos around MSPs is higher
than that of solar neighbourhood. 

Here, we estimate the residual of the time of arrival of a MSP due to 
the gravitational attraction from its
nearest neighbor, following \cite{Seto2007}.
The residual due to
the constant acceleration $a$ in the period of $t$ is given by
\begin{equation}
  \Delta T =\frac{ at^2}{2c},
\end{equation}
where $c$ is the speed of light, and $a$ is given by
\begin{equation}
  a = -\frac{GM}{R^2},
\end{equation}
where $G$, $M$ and $R$ are the gravitational constant, the mass of the
microhalo, and the distance between the microhalo and the MSP. So we
have
\begin{equation}
  \Delta T = 10.8 \left(\frac{R}{2\times 10^4{\rm AU}}\right)^{-2}
                   \left(\frac{M}{10^{-6} M_{\odot}}\right)
                   \left(\frac{t}{\rm 20yr} \right)^{2} {\rm ns}.
\end{equation}
If we assume that a microhalo lost 90\% of mass by encounters with stars, 
the residual is about $\sim 2 {\rm ns}$ in twenty years. The
change of the distance in ten years is around 400AU.
Since the change in the acceleration is proportional to the third power
of the distance, one in ten MSPs would show the change in the
acceleration ten times larger than that of a typical MSP, which should
be detectable with PPTA \citep{Manchester2008} after modest improvement
in the timing accuracy. 

Note that the mass loss of two weak encounters can be larger than that
of a strong encounter \citep{Angus2007}.  For simplicity, we used the
mass loss due to a strong encounter as estimated in subsection
\ref{sec:encounter}.
However, the mass loss of microhalos in the opossite direction to the
galactic center is less than that near Sun. MSPs in the direction should
show larger timing residual.

\subsection{Summary}
We found that earth-mass microhalos have steep central density cusps.
Their central regions is not disrupted by tidal forces from the parent
galaxies or stars and they survive to the present time.  Our result is
different from the recent claims that small-scale structure have a
negligible impact on dark matter detectability \citep{Springel2008}.
They considerably underestimate annihilation signals because they
assumed the density profile for the smallest microhalos was
shallower than that we found with high resolution simulations.

\acknowledgements
We are grateful to Veniamin Berezinsky for helpful discussions.
Numerical computations were carried out on Cray XT4 at Center
for Computational Astrophysics, CfCA, of National Astronomical
Observatory of Japan.  T.I. is financially supported by Research
Fellowship of the Japan Society for the Promotion of Science (JSPS)
for Young Scientists.
This research is partially supported by the
Special Coordination Fund for Promoting Science and Technology
(GRAPE-DR project), Ministry of Education, Culture, Sports, Science
and Technology, Japan.

\end{document}